# Modeling electric field sensitive scanning probe measurements for a tip of arbitrary shape

I. Kuljanishvili[a], Subhasish Chakraborty[a], I .J. Maasilta[a]*, S. H. Tessmer[a], and M. R. Melloch[b]

[a] *Department of Physics and Astronomy, Michigan State University, East Lansing, MI 48824*

[b] *Department of Electrical Engineering, Purdue University, West Lafayette, Indiana 47907*

**Abstract**

We present a numerical method to model electric field sensitive scanning probe microscopy measurements which allows for a tip of arbtrary shape and invokes image charges to exactly account for a sample dielectric overlayer. The method is applied to calculate the spatial resolution of a subsurface charge accumulation imaging system, achieving reasonable agreement with experiment.

## 1. Introduction

The tip geometry can be a critical factor for electric-field-sensitive scanning probe microscopies, such as scanning capacitance microscopy [1-3], scanning single-electron transistor microscopy [4,5], charged-probe atomic force microscopy [6,7], and subsurface charge accumulation (SCA) imaging [8,9]. The samples of interest typically consist of conducting layers or nanostructures buried beneath a dielectric. To interpret the data and the influence of the tip geometry, researchers often rely on modeling in which the tip is taken to have an ideal shape such as a perfect cone or sphere [10]. However in many cases the tip may be better described by a less regular shape. For example, the end of the tip may break due to contact with the surface resulting in a truncated cone. Moreover, the tip may be bent or mounted in a non-perpendicular angle so that any model that relies on cylindrical symmetry would be inappropriate. We have developed a straightforward numerical method that can conveniently model a tip of *arbitrary* shape.

Our method is based on a boundary element approach – inspired by the numerical method used to model the charging patterns within a GaAs-AlGaAs two-dimensional electron layer probed with the SCA technique [11]. As shown in Figure 1(a), we approximate continuous tip and sample electrodes as arrays of point-like conducting elements. By considering the Coulomb interaction among all the elements and using image charges to account exactly for a dielectric overlayer, we directly solve for the complete self-capacitances and mutual capacitance for the tip-sample system. In this paper we introduce the method and demonstrate its utility by calculating the mutual capacitance of a realistic tip and sample used for an SCA measurement. We then compare the resulting modeled spatial resolution to an experimental measurement.

## 2. Electrostatic framework

Following basic electrostatics [12], for a system of $n$ conductors each with potential $V_i$ and charge $Q_i$, the voltage is a linear function of charge:

$$V_i = \sum_{k=1}^{n} A_{ik} Q_k , \quad (1)$$

where the elements $A_{ik}$ represent a potential matrix $\hat{A}$, and give the voltage of point $i$ due to unit charge at point $k$ and all other charges equal to zero. The matrix is symmetric: $A_{ik} = A_{ki}$. In our typical calculations, roughly 25% of the $n$ conductors correspond to tip points and 75% correspond to sample points. Inverting Eq.1 yields an expression for the charge on each conductor, which depends highly on the surrounding conductors and dielectric medium:

$$Q_i = \sum_{k=1}^{n} C_{ik} V_k, \quad \hat{C} = \hat{A}^{-1}, \quad (2)$$

where the $C_{ik}$ are the coefficients that make up the capacitance matrix $\hat{C}$.

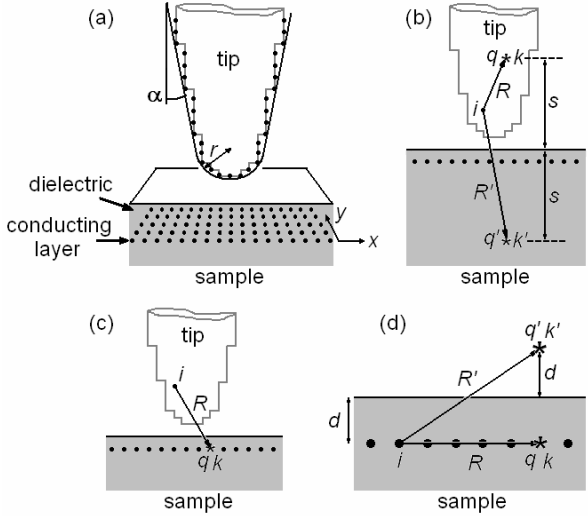

Fig. 1. (a) Schematic picture of the boundary element approach, showing the tip-sample geometry used to model subsurface charge accumulation measurements. The tip is represented by discrete conducting points arranged to form a steep pyramid of half-angle $\alpha$. The apex is curved with radius $r$. The sample conductor is represented with points arranged in a two-dimensional square grid imbedded in a dielectric. (b) Schematic of the calculation of potential matrix element $A_{ik}$ for *Case 1*, for which both $i$ and $k$ are points in the tip and $k$ is a distance $s$ from the dielectric surface. The image charge at the symmetric position with respect to the vacuum-dielectric interface $k'$ accounts exactly for the modification in potential due to the dielectric layer. $R$ and $R'$ are the distances from $i$ to $k$ and $k'$, respectively. (c) Schematic for *Case 2*, for which $i$ is in the tip and $k$ in the sample. Here the dielectric layer is accounted for using an image charge at the location of the actual charge. (d) Schematic for *Case 3*, for which for which both $i$ and $k$ are points in the sample a distance $d$ below the dielectric. In analogy to *Case 1*, the dielectric layer is accounted for by invoking an image charge at the symmetric position $k'$.

Our scheme follows directly from equations (1) and (2). We first construct $\hat{A}$ by considering the Coulomb interaction among all pairs of points and invoking image charges to account for bound charges at the dielectric surface, as described below. $\hat{C}$ is then found by inverting $\hat{A}$. The tip-sample mutual capacitance is a quantity of considerable importance for electric-field-based microscopies. Once we have solved for all $C_{ik}$, we can use Eq. 2 to calculate the mutual capacitance. Clearly, the calculation must treat tip points differently from sample points. In constructing $\hat{A}$, three cases arise depending on the identity of each of the pair of points, as detailed in the following subsections.

## 2.1. Case 1: tip-tip

Here we consider both $i$ and $k$ to be conductors in the tip. The potential at $i$ due to unit charge at $k$ must also consider the image charge at $k'$ to account for the dielectric, as shown in Fig. 1(b). In this case the image charge solution [12] is

$$q' = -\left(\frac{\kappa - 1}{\kappa + 1}\right) q,$$

where $\kappa$ is dielectric constant, $q$ is a unit charge located at point $k$ and $q'$ is it's image located at the symmetric position about the vacuum-dielectric interface, $k'$. The potential matrix element then becomes

$$A_{ik} = \frac{1}{R} - \left(\frac{\kappa - 1}{\kappa + 1}\right)\frac{1}{R'},$$

where $R$ and $R'$ are distances from point $i$ to points $k$ and $k'$ respectively.

## 2.2. Case 2: tip-sample

Similarly we calculate the coefficients of potential for point $i$ in the tip and point $k$ in the sample. In this case, point $k$ is immersed in the dielectric and the image charge solution results in a simple reduction of the potential:

$$A_{ik} = \left(\frac{2}{\kappa + 1}\right)\frac{1}{R},$$

where $R$ is the distance between points $i$ and $k$, as shown in Fig 1(c).

*2.3. Case 3: sample-sample*

Lastly, if both *i* and *k* are conductors in the sample, the dielectric surface is accounted for with an image charge term analogous to *Case 1*.

$$A_{ik} = \frac{1}{\kappa R} + \frac{1}{\kappa}\left(\frac{\kappa-1}{\kappa+1}\right)\frac{1}{R'},$$

where $R$ is the distance from conductor $i$ to $k$ in the sample, and $R'$ is the distance from conductor $i$ to $k'$, located at the symmetric position about the vacuum-dielectric interface, as shown in Fig.1(d).

## 3. Results and discussion

*3.1. SCA measurements*

We have employed subsurface charge accumulation imaging to study the local structure of the interior of a two-dimensional electron system embedded within a GaAs/AlGaAs heterostructure. For the data we will model here, the sample contained a 2D layer 60 nm below the exposed surface; the dielectric constant of the overlayer is approximately $\kappa = 12.2$. The tip was a chemically-etched PtIr wire with a half angle $\alpha \approx 8°$ for the conical section, and a nominal apex radius of curvature of $r = 50$nm [13]. Below the 2D layer the sample contained a heavily doped metallic substrate (3D) separated from the 2D layer by a tunneling barrier [14,15]. An *ac* excitation voltage (20 kHz, 8mV rms) applied to the 3D substrate causes electrons to tunnel from the 3D to the 2D. The charge that enters the 2D layer induces *ac* image charge on the apex of the tip, which is positioned a few nm from the surface. We measure this capacitively induced image charge $Q_{tip}$ using a cryogenic high electron mobility transistor attached directly to the tip.

The local measured charge induced on the tip can be expressed as

$$Q_{tip}(x_0, y_0) = \int \phi(x,y) c_{mut}(x-x_0, y-y_0) dx dy ,\quad(3)$$

where $\phi(x, y)$ is the effective local ac potential in the 2D layer and $c_{mut}(x, y)$ is the tip-2D mutual capacitance per unit area of the 2D layer [11]; $c_{mut}(x, y)$ is a crucial function which sets the spatial resolution of the technique. Below we describe the application of our method to calculate this function.

*3.2. Mutual capacitance calculation*

We modeled the tip as a set of point conductors forming a steep pyramid terminated with a rounded apex, as shown in Fig. 1(a), with $\alpha = 8°$ and $r = 50$ nm. The apex was positioned directly above the center of the sample a distance of 5 nm from the surface. The sample model consisted of a square array of grid points to form the 2D layer, fixed at a position of $d$=60 nm below a $\kappa = 12.2$ dielectric layer. We used a spacing of 20 nm between adjacent points for both the points that represent the tip and the grid points that form the 2D layer to approximate continuous conductors. Of course, neither the tip nor sample can be infinite in extent; we took the total tip height and sample width to be 0.52 μm and 1.06 μm, respectively. With respect to length scales ~100 nm, these sizes are sufficiently large so that edge effects do not contribute significantly to the charging near the center of the sample and near the tip's apex. The total number of conductors (tip+sample) in the calculation was 3,746.

After solving for $\hat{C}=\hat{A}^{-1}$, we can find $c_{mut}(x, y)$ by calculating the charge on each grid point of the 2D layer per unit potential difference applied between the tip and sample. This is most conveniently accomplished by considering the tip to be at potential $V$ with the sample grounded. The summation in Eq. (2) then involves only points in the tip. We can further rewrite Eq. 2 by dividing each term by voltage and area:

$$c_{mut}(x_i, y_i) = Q_i/(Va) = \sum_{k}^{tip} C_{ik}/a ,$$

where $(x_i, y_i)$ refer to the coordinates of point $i$ in the sample, and $a$ is sample area per grid point.

Figure 2 shows the resulting calculated mutual capacitance function $c_{mut}(x, y)$. We see that it is a bell-shaped function peaked directly below the tip's apex at the center of the sample. The breadth of this peak determines the spatial resolution of the measurement, which we characterize using the half-width at half-maximum, $w$=92 nm. To check the errors introduced by the discrete grid spacings and truncated tip and sample, we performed similar calculations using spherically shaped tips with no dielectric layer, and then compared the calculations directly to the analytical solutions. We estimate the plotted $c_{mut}(x, y)$ function is accurate to within 10%, with the exception of points near the edges.

*3.3. Comparison to measured data*

To compare these calculations to measured SCA images, it would be instructive to select an image exhibiting an especially sharp feature so that we can assume the true physical extent of the feature is much smaller than $w$. In that case, the measured width would be mostly determined by the experimental resolution. Figure 3 shows such a comparison. The measurement was performed at 1º K in a top loading helium-3 cryostat. A depleting voltage of -0.23 V was applied to the tip which serves to significantly enhance the effect of density variations to produce contrast in the measurement [15]. The imaged features simply reflect sample disorder. To calculate the model charging profile we assume that the effective potential $\phi(x, y)$ varies as an arbitrarily sharp step function along the direction of the white line. The step function is smeared out by convolving it with $c_{mut}(x,y)$ following Eq. 3. As shown in Fig. 3(b), the resulting curve reasonably approximates the experimental spatial resolution with no adjustable parameters.

The common picture for the spatial resolution of electric field sensitive scanning probe measurements holds that the radius of curvature of the tip and the tip-sample conductor separation set the resolution length scale. Because the radius of curvature is the smaller length for our modeled system, we can estimate roughly that the length scale should equal 65 nm, the depth of the 2D layer ($d$) plus the additional tip to surface distance. This simple estimate is supported by calculations of Eriksson *et al.* based on a numerical routine that solves the Laplace equation in an axially symmetric geometry. [16] For a similar tip-sample system with the tip in direct contact with the dielectric surface, the calculations showed that the half-width at half-maximum of $c_{mut}(x, y)$ was approximately equal to the depth of the 2D layer, $w \approx 1.0d$. (The agreement between $w$ and $d$ was to within 5% as long as the area of contact was less than about 25 nm.) In contrast both our calculation and our experimental SCA measurements, in which the tip is not directly touching the surface, show a significantly larger spatial resolution length scale of $w=1.5d$.

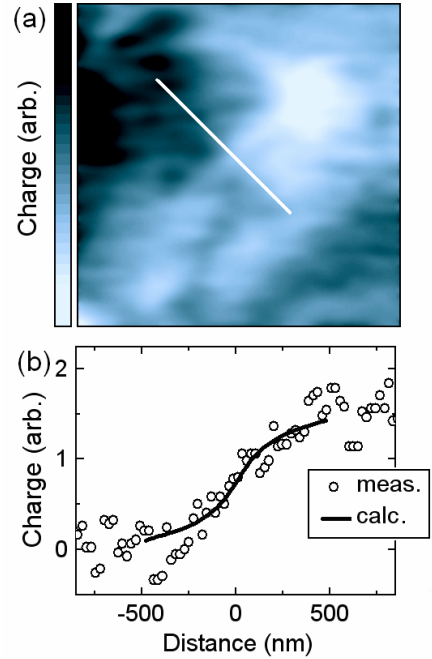

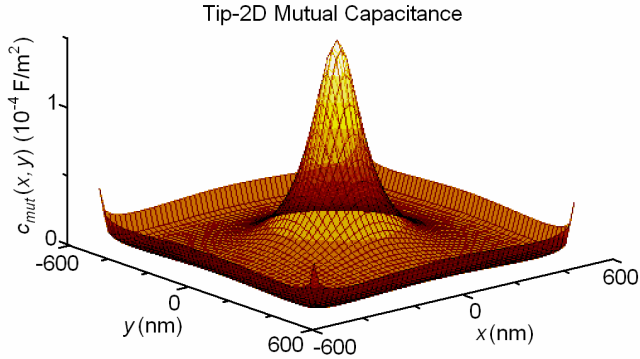

Fig. 2. Calculated mutual capacitance function $c(x,y)$ showing the characteristic bell shape. The calculation used the tip geometry shown if Fig. 1(a) with a half angle $\alpha =8°$, radius $r=50$ nm, total height of 520 nm, and tip-surface separation of 5 nm. The modeled sample was 1.06 x 1.06 μm, with a dielectric overlayer of $d=60$ nm. For both the tip and sample the spacing between adjacent grid points was 20 nm. The half width at half maximum is 92 nm, which determines the spatial resolution of our technique.

Fig. 3. (a) A 3 x 3 μm subsurface charge accumulation image of a GaAs-AlGaAs two-dimensional electron layer formed in a GaAs-AlGaAs heterostructure. The contribution of surface topography has been subtracted so that the displayed signal arises solely from disorder within the two-dimensional layer. The image is filtered to remove nanometer scale noise. For comparison to modeling we focus on the data indicated by the white line; here the signal increases abruptly, representative of the sharpest features imaged by the technique. (b) Cross section of the image along the white line, compared directly to our model calculation as described in the text. In this case no filtering was applied to the measured data, and no adjustable parameters were used in the calculation to achieve the fit.

## 4. Conclusion

In conclusion, we have developed a numerical method to model electric field sensitive scanning probe microscopy measurements. It is a boundary element approach that uses image charges to exactly account for a sample dielectric overlayer. The elements can be arranged in three dimensions, with no restrictions on the symmetry of the tip geometry – hence it is straightforward to model a tip of arbitrary shape. We have applied method to calculate the expected spatial resolution of a subsurface charge accumulation imaging system. The model predicts a spatial resolution length scale 50% greater than the thickness of the dielectric overlayer, in reasonable agreement with the sharpest features seen in the images.

## Acknowledgments

We thank L.S. Levitov for crucial discussions and input in developing the modeling method. This work was supported by the National Science Foundation grant nos. DMR00-75230 and DMR03-05461. SHT acknowledges support of the Alfred P. Sloan Foundation.